# A giant comet-like cloud of hydrogen escaping the warm Neptune-mass exoplanet GJ 436b

David Ehrenreich[1], Vincent Bourrier[1], Peter J. Wheatley[2], Alain Lecavelier des Etangs[3,4], Guillaume Hébrard[3,4,5], Stéphane Udry[1], Xavier Bonfils[6,7], Xavier Delfosse[6,7], Jean-Michel Désert[8], David K. Sing[9] & Alfred Vidal-Madjar[3,4]

**Exoplanets orbiting close to their parent stars could lose some fraction of their atmospheres because of the extreme irradiation[1–6]. Atmospheric mass loss primarily affects low-mass exoplanets, leading to suggest that hot rocky planets[7–9] might have begun as Neptune-like[10–16], but subsequently lost all of their atmospheres; however, no confident measurements have hitherto been available. The signature of this loss could be observed in the ultraviolet spectrum, when the planet and its escaping atmosphere transit the star, giving rise to deeper and longer transit signatures than in the optical spectrum[17]. Here we report that in the ultraviolet the Neptune-mass exoplanet GJ 436b (also known as Gliese 436b) has transit depths of $56.3 \pm 3.5\%$ ($1\sigma$), far beyond the 0.69% optical transit depth. The ultraviolet transits repeatedly start ~2 h before, and end >3 h after the ~1 h optical transit, which is substantially different from one previous claim[6] (based on an inaccurate ephemeris). We infer from this that the planet is surrounded and trailed by a large exospheric cloud composed mainly of hydrogen atoms. We estimate a mass-loss rate in the range of $\sim 10^8$–$10^9$ g s$^{-1}$, which today is far too small to deplete the atmosphere of a Neptune-like planet in the lifetime of the parent star, but would have been much greater in the past.**

Three transits of GJ 436b, which occur every 2.64 days, were observed on 7 December 2012 (ref. 6) (visit 1), 18 June 2013 (visit 2) and 23 June 2014 (visit 3) with the Space Telescope Imaging Spectrograph (STIS) on board the Hubble Space Telescope (HST). A stellar spectrum acquired using similar settings in January 2010 (ref. 17) (visit 0) was retrieved from the archive for comparison purposes. HST data in visits 2 and 3 were complemented with simultaneous Chandra X-ray observations. The HST data consist of time-tagged, far-ultraviolet spectra obtained with a grating dispersing light over the 1,195–1,248 Å domain, with a spectral resolution of ~20 km s$^{-1}$ at 1,215.6 Å (the Lyman-$\alpha$ transition of atomic hydrogen, H I). Exposure times of 1,500 s to 2,900 s were used to observe the star for four successive HST orbits during each visit. Each HST orbit lasts for 96 min, during which GJ 436 is visible for 56 min before being occulted by the Earth, yielding 40 min gaps in the data.

[1]Observatoire de l'Université de Genève, 51 chemin des Maillettes, 1290 Versoix, Switzerland. [2]Department of Physics, University of Warwick, Coventry CV4 7AL, UK. [3]CNRS, UMR 7095, Institut d'Astrophysique de Paris, 98 bis boulevard Arago, 75014 Paris, France. [4]Sorbonnes Universités, UPMC Univ Paris 6, UMR 7095, Institut d'Astrophysique de Paris, 98 bis boulevard Arago, 75014 Paris, France. [5]Observatoire de Haute-Provence, CNRS & OAMP, 04870 Saint-Michel-l'Observatoire, France. [6]Univ. Grenoble Alpes, IPAG, F-38000 Grenoble, France. [7]CNRS, IPAG, F-38000 Grenoble, France. [8]CASA, Department of Astrophysical & Planetary Sciences, University of Colorado, 389-UCB, Boulder, Colorado 80309, USA. [9]Astrophysics Group, School of Physics, University of Exeter, Stocker Road, Exeter EX4 4QL, UK.

The most prominent spectral feature is the H I Lyman-α emission (Fig. 1). Absorption in the blue wing of this line has been reported in other systems, during transits of hot Jupiters. This is interpreted by the presence of escaping hydrogen exospheres surrounding giant planets[1,5,18–20]. Tentative evidence that the Neptune-mass planet GJ 436b possesses such an extended atmosphere was drawn from visit 1 data despite the signal being observed after one optical transit[6], raising questions on its planetary origin. Visits 2 and 3 were carried out to determine the signal nature.

We performed a careful analysis to check for the existence of instrumental systematics in the data and correct for them (see Methods). Large variations are detected over a part of the Lyman-α line at times corresponding to the optical transit, which cannot be explained by any known instrumental effects. The most notable absorption occurs in the blue wing of the line for radial velocities between −120 km s$^{-1}$ to −40 km s$^{-1}$, ~2 h before the optical mid-transit time ('pre-transit'), during the optical transit ('in-transit') and ~1 h after the optical mid-transit time and beyond ('post-transit') in the three visits (Fig. 1 and Extended Data Fig. 1). In-transit, over half the stellar disc (56.2 ± 3.6%, 1$\sigma$) is eclipsed (Fig. 2a). This is far deeper than any ultraviolet transits of hot Jupiters and significantly (~9$\sigma$) deeper than the 22.9 ± 3.9% (1$\sigma$) post-transit signal previously reported in visit 1 data[6]. The ultraviolet transit also starts much earlier (~2.7 h) than claimed previously; the difference is mainly due to our finding of a pre-transit absorption and updated transit ephemeris (see Methods). Visit 0 data[17] acquired ~3 years before visit 1 and ~30 h after transit show that the out-of-transit variability is small compared with the blue-shifted signature. This gives us confidence that the true out-of-transit baseline is measured in visits 2 and 3, whereas it is missing in visit 1. By contrast, the flux remains stable over the whole red-shifted wing of the line (Fig. 2b). The decrease of the red-wing flux seen[6] during the post-transit phases of visit 1 is not reproduced during visits 2 and 3. The mean post-transit red-shifted signal is compatible with no detection at the 3$\sigma$ level.

Our combined analysis of X-ray and ultraviolet data (see Methods) shows that stellar magnetic activity cannot explain the observed decrease at Lyman-α. We propose that the asymmetric absorption is caused by the passage of a huge hydrogen cloud, surrounding and trailing the planet (Fig. 3). The planetary atmosphere is an obvious source for this hydrogen. To produce this extinction signature, we estimate that an ellipsoidal, optically thick cloud of neutral hydrogen should have a projected extension in the plan of the sky of ~12 stellar radii ($R_* \approx 0.44$ $R_\odot$) along the orbital path of the planet and ~2.5 $R_*$ in the cross direction, well beyond the planet Roche lobe radius (0.37 $R_*$). Since GJ 436b grazes the stellar disc during transit, we surmise that a central transit would have totally eclipsed the star. This could happen in the case of other red dwarfs exhibiting central transits from planets similar to GJ 436b. Future ultraviolet observations of systems similar to GJ 436 could potentially reveal total Lyman-α eclipses.

The radial velocity interval of the absorption signal constrains the dynamics of the hydrogen atoms and the three-dimensional structure of the exospheric cloud. The whole velocity range is in excess of the planet escape velocity (~26 km s$^{-1}$ at the planet surface), consistent with gas escaping from the planet. The acceleration mechanism of hydrogen atoms escaping from highly irradiated hot Jupiters is debated: after escaping the planets with initial velocities dominated by the orbital velocity (~100 km s$^{-1}$ for GJ 436b in the host star reference frame), atoms are submitted to the stellar radiation pressure, interact with the stellar wind and are eventually ionized by stellar extreme ultraviolet (EUV; 10–91.2 nm) radiation. For strong lines such as Lyman-α, radiation pressure can overcome the stellar gravity, repelling the escaping atoms towards the observer and producing a blue-shifted signature. In one hot Jupiter (HD 189733b), the absorption observed at large blue-shifts is best explained by charge exchange interaction with the stellar wind, creating energetic neutral atoms with large blue-shifted radial velocities[19–22]. In other cases[1,18,23], radiation pressure alone explains the observed radial velocities of the escaping gas.

We ran a three-dimensional numerical simulation of atmospheric escape[20] to understand the origin of the absorption signature observed at GJ 436b. The line profile corrected from interstellar absorption[17,24] is used to calculate the stellar radiation pressure on hydrogen atoms. These are released isotropically from the Roche lobe limit of GJ 436b. The calculation takes the orbital eccentricity of GJ 436b ($e = 0.15$) into account. The main parameters of the atmospheric escape model are the mass-loss rate $\dot{m}$ of hydrogen lost by the planet and the photo-ionization rate $\alpha_{EUV}$ of hydrogen atoms. The model computes the structure of the escaping gas cloud, its radial velocity absorption signature, and the integrated light curve. Adjusting for the spectra at different phases (Extended Data Fig. 2), we determine that a family of models, with parameters in the $1\sigma$ ranges of $\dot{m} \approx 10^8$–$10^9$ g s$^{-1}$ and $\alpha_{EUV} \approx 8 \times 10^{-7}$–$3 \times 10^{-6}$ s$^{-1}$ (implying neutral atom lifetimes of ~4–18 min at the distance of the planet), provide good fits to the data, yielding reduced $\chi^2$ values close to unity for 544 degrees of freedom. A synthetic light curve is plotted in Fig. 2. The model correctly matches the ~2 h early ultraviolet transit ingress observed with respect to the optical transit, as well as the transit depth in the correct range of velocities. It provides a good match to the reanalysed visit 1 data. It furthermore predicts that the ultraviolet transit could last up to ~20 h after the optical transit, due to the extended hydrogen tail of the exospheric cloud. More ultraviolet observations will be needed to verify this prediction.

According to our simulation, the stellar radiation pressure counterbalances ≲70% of the star's gravity pull on escaping atoms, which is much less than in hot Jupiter systems, in which radiation pressure takes over stellar gravity by factors of 3 to 5 (ref. 20). The low stellar radiation pressure at GJ 436b allows the formation of a large coma and tail of escaping atoms, co-moving with the planet although not gravitationally bounded to it (Fig. 4).

Atmospheric escape is involved in the possible loss of a whole population of irradiated exoplanets[11,13,15,16]. The average mass-loss rate of ~$5 \times 10^8$ g s$^{-1}$ at GJ 436b means that the planet loses ~0.1% of its atmosphere per billion years (assuming it accounts for 10% of the planet mass, like Neptune). This rate requires ~1% efficiency in the conversion of input X and EUV energy (estimated in Methods) into mass loss[15]. In the past, an M dwarf such as GJ 436 would have been more active and the planet could have received ≲100 times more X-ray and EUV irradiation over ~1 Gyr[25], resulting in a possible loss of ≲10% of its atmosphere during the first billion years. This planet thus stands on the edge of considerable mass loss, leading us to surmise that closer-in Neptunes could have evolved more dramatically because of atmospheric escape.

This $16\sigma$ detection implies that large atmospheric signals from comet-like exospheres around moderately irradiated, low-mass planets could be retrieved in the ultraviolet. Over ~10,000 systems similar to GJ 436 will be discovered by upcoming transit surveys carried out from the ground and from space—for example, with the K2, CHEOPS, TESS and PLATO missions.

**Acknowledgements** This work is based on observations made with the NASA/ESA Hubble Space Telescope, obtained at the Space Telescope Science Institute, which is operated by the Association of Universities for Research in Astronomy, Inc., under NASA contract NAS 5-26555. These observations are associated with programmes #11817, #12034 and #12965. The scientific results reported in this article are based on observations made by the Chandra X-ray Observatory. This work was carried out in the framework of the National Centre for Competence in Research 'PlanetS' supported by the Swiss National Science Foundation (SNSF). D.E., V.B. and S.U. acknowledge the financial support of the SNSF. V.B., A.L.d.E., X.B. and X.D. acknowledge the support of CNES, the French Agence Nationale de la Recherche (ANR) under program ANR-12-BS05-0012 'Exo-Atmos', the Fondation Simone et Cino Del Duca, and the European Research Council (ERC) under ERC Grant Agreement no. 337591-ExTrA.

**Author Contributions** D.E. proposed and led the HST–Chandra joint observation programme, supervised data reduction and analysis, interpreted the results and wrote the paper. V.B. performed data reduction and analysis, and computer simulations to interpret the results. P.J.W. set up the Chandra X-ray observations, reduced, analysed and interpreted the X-ray data. A.L.d.E. co-designed the simulation programme with V.B. and provided computing resources to run the simulations. A.L.d.E. and G.H. contributed to the observation programme, data analysis and interpretation. S.U., X.B., X.D., J.-M.D., D.K.S. and A.V.-M. contributed to the observation programme and interpretation. All authors discussed the results and commented on the manuscript.

**Author Information** Reprints and permissions information is available at www.nature.com/reprints. The authors declare no competing financial interests. Readers are welcome to comment on the online version of the paper. Correspondence and requests for materials should be addressed to D.E. (david.ehrenreich@unige.ch).

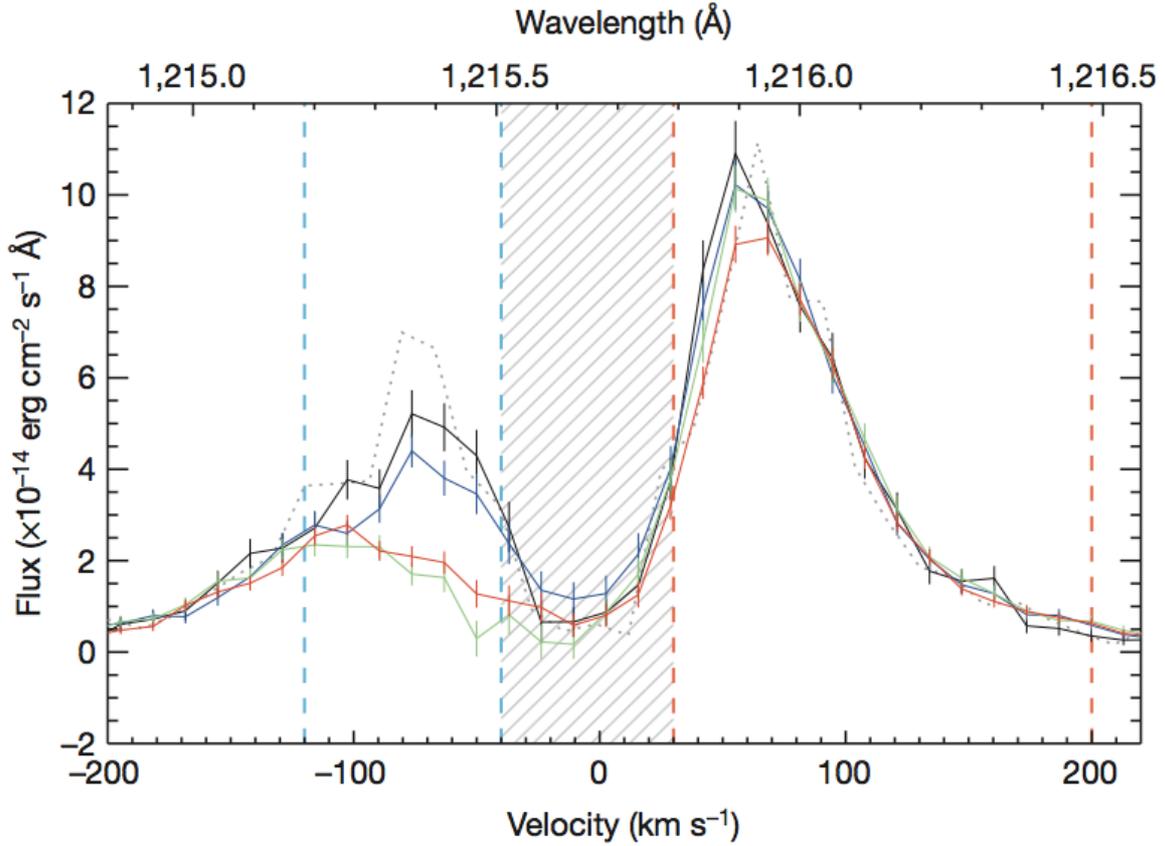

**Figure 1 | Evolution of the hydrogen Lyman-α emission line of GJ 436.** The line has been averaged over out-of-transit (black), pre-transit (blue), in-transit (green) and post-transit (red) observations from individual spectra (see Extended Data Fig. 1). The $1\sigma$ uncertainties have been propagated accordingly from the errors calculated by the STIS data reduction pipeline. The line profile from visit 0 is shown for comparison with a dotted grey line. The line core (hatched region) cannot be observed from Earth because of the interstellar medium absorption along the line of sight. Absorption signals are measured over the interval $[-120,-40]$ km s$^{-1}$ (blue dashed lines) and compared to a control measure over the interval $[+30,+200]$ km s$^{-1}$ (red dashed lines). The velocity scale is heliocentric.

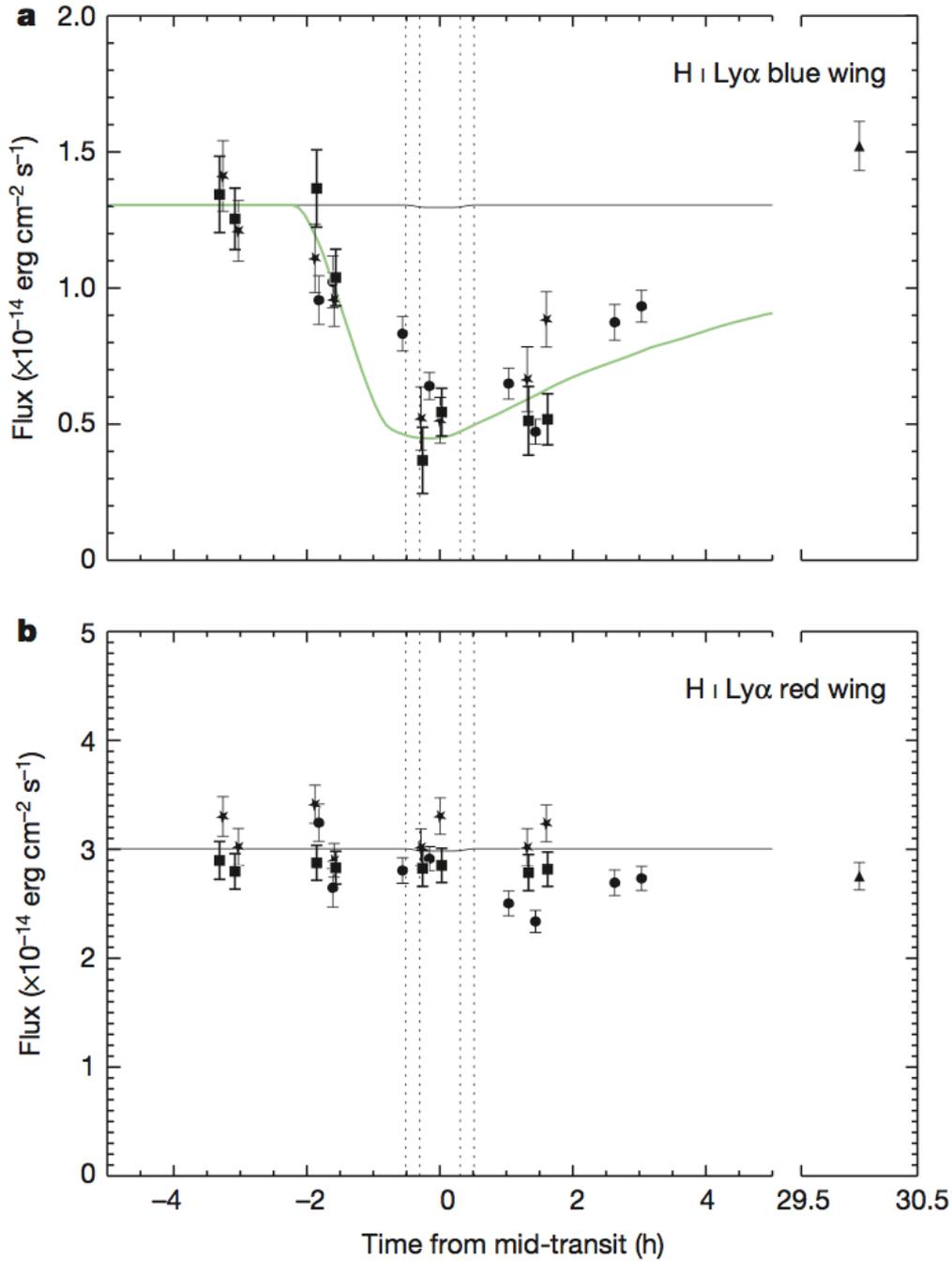

**Figure 2 | Lyman-α transit light curves of GJ 436b. a, b**, Data are from visit 1 (circles), visit 2 (stars), visit 3 (squares) and visit 0 (triangles). All uncertainties are 1σ. **a**, The Lyman-α (Lyα) line is integrated over [−120,−40] km s$^{-1}$ and shows mean absorption signals with respect to the out-of-transit flux of 17.6 ± 5.2% (pre-transit), 56.2 ± 3.6% (in-transit) and 47.2 ± 4.1% (post-transit). **b**, The line is integrated over [+30,+200] km s$^{-1}$ and shows no significant absorption signals: 0.7 ± 3.6% (pre-transit), 1.7 ± 3.5% (in-transit) and 8.0 ± 3.1% (post-transit). With a depth of 0.69%, the optical transit (thin black lines in **a** and **b**) is barely seen at this scale between its contact points (dotted lines in **a** and **b**). A synthetic light curve (green) calculated from the three-dimensional numerical simulation[20] is overplotted on the data in **a**.

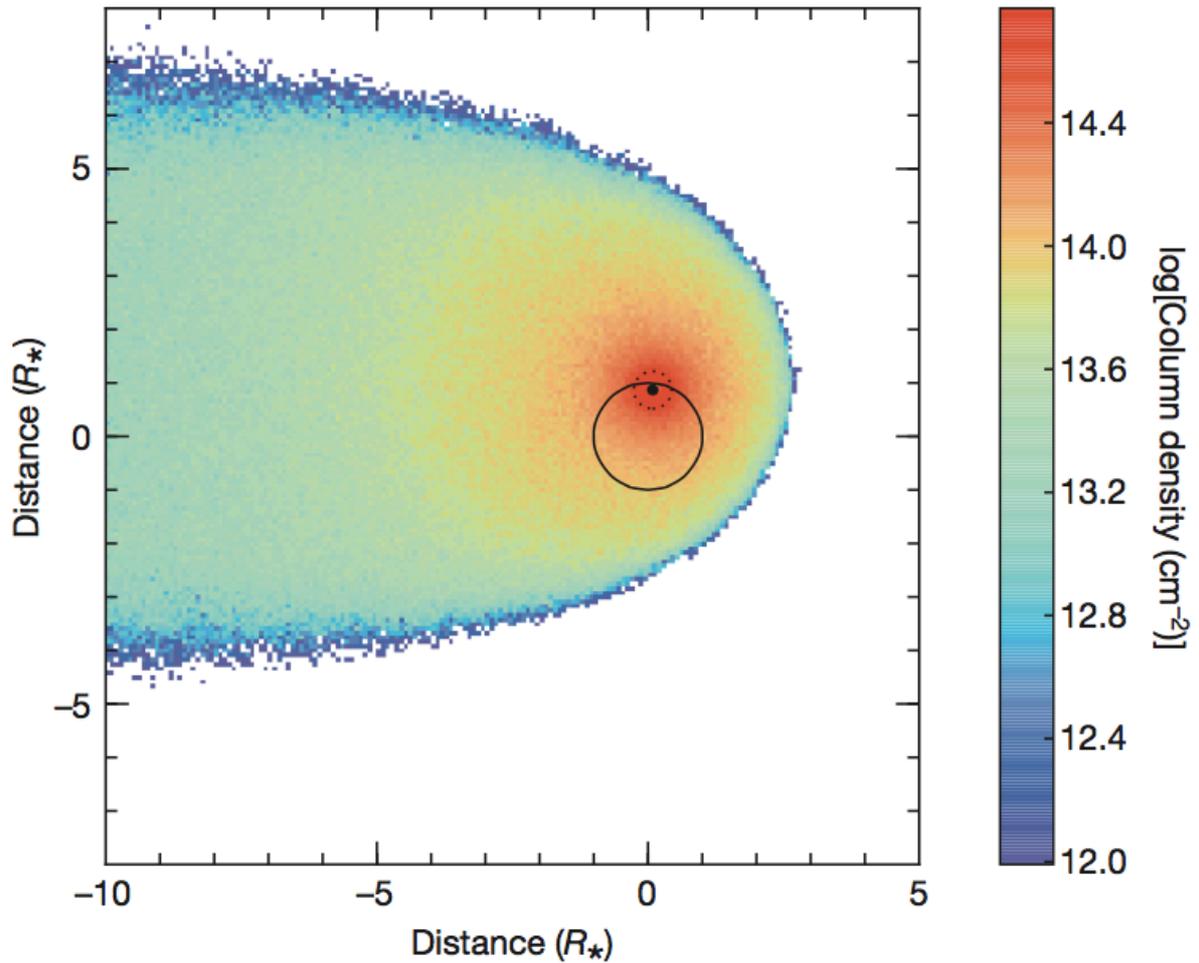

**Figure 3 | Particle simulation showing the comet-like exospheric cloud transiting the star, as seen from Earth.** GJ 436b is the small black dot represented at mid-transit at 0.8521 $R_*$ (ref. 26) from the centre of the star, which is figured by the largest black circle. The dotted circle around the planet represents its equivalent Roche radius. The colour of simulation particles denotes the column density of the cloud. The transit of this simulated cloud gives rise to absorption over the blue wing of the Lyman-α line as shown spectrally in Extended Data Fig. 2 and by the synthetic light curve in Fig. 2a.

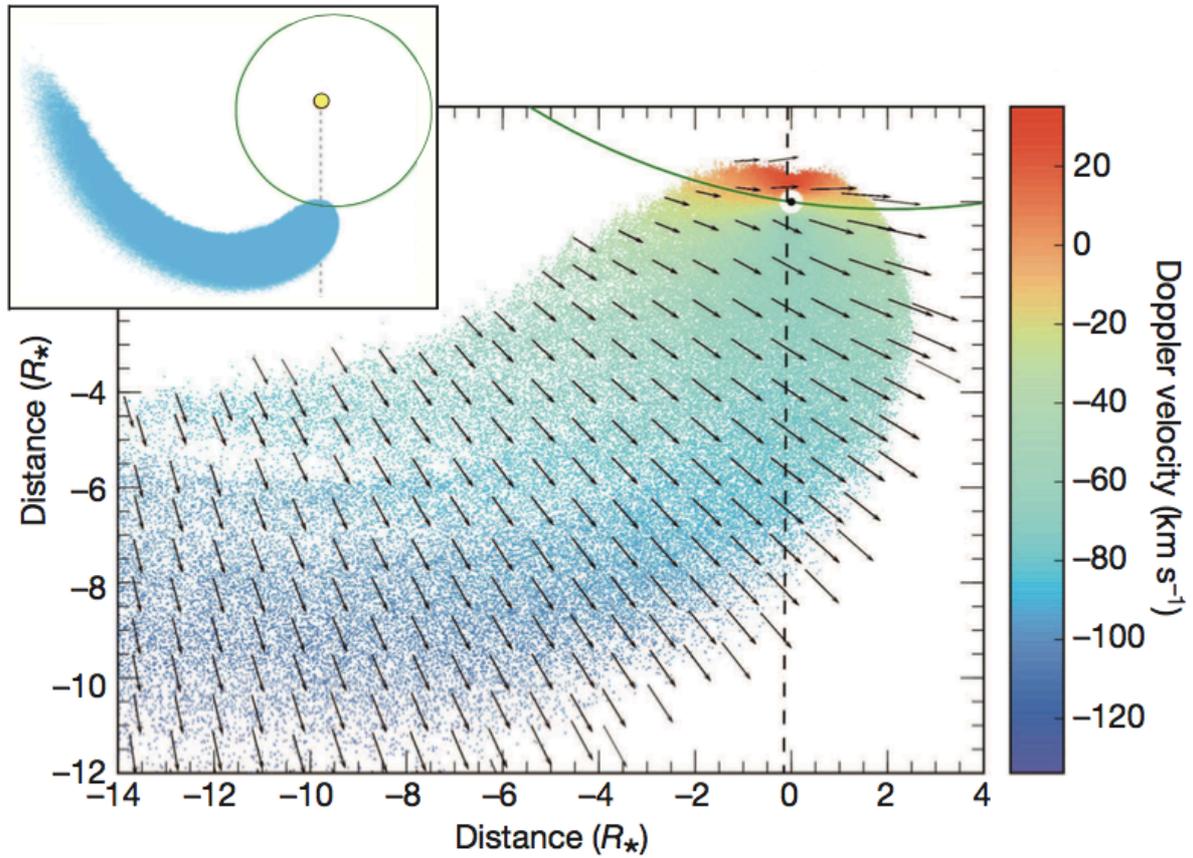

**Figure 4 | Polar view of three-dimensional simulation representing a slice of the comet-like cloud coplanar with the line of sight.** Hydrogen atom velocity and direction in the rest frame of the star are represented by arrows. Particles are colour-coded as a function of their projected velocities on the line of sight (the dashed vertical line). Inset, zoom out of this image to the full spatial extent of the exospheric cloud (in blue). The planet orbit is shown to scale with the green ellipse and the star is represented with the yellow circle.

**METHODS**

**Observation log**

This work is mainly based on observations acquired in the framework of the joint HST–Chandra programme "Properties and dynamics of the upper atmosphere of the hot-Neptune GJ 436b" (GO number 12965; principal investigator: D.E.). The programme consisted of three HST/STIS transit observations, with contemporaneous coverage of the pre-transit phases with Chandra/ACIS-S. During the first joint HST–Chandra visit, initially scheduled on 16 February 2013, we only obtained Chandra data as a last-minute mission scheduling change prevented the HST observation. It was rescheduled, together with a new Chandra observation, on 18 April 2013. However, no useful HST data could again be obtained on this date because of a target pointing problem. Meanwhile, the simultaneous Chandra observation was correctly performed. The remaining two HST–Chandra visits were smoothly carried out on 18 June 2013 and 23 June 2014.

We complement our new data sets with archive HST/STIS observations of GJ 436: (1) a short observation of the star obtained ~30 h after the planetary transit on 5 January 2010 (HST programme GO/DD number 11817 entitled "Detecting the upper atmosphere of a transiting hot Neptune: a feasibility study"; principal investigator: D.E.[17]); and (2) a transit observation obtained on 7 December 2012 (HST programme GTO number 12034 entitled "Brown dwarf activity part 2"; principal investigator: J.C. Green). The 7 December 2012 transit observation was previously published[6] and is discussed later. All HST and Chandra data are listed in Extended Data Table 1.

Similarly, we add archive X-ray data of GJ 436 to our new Chandra observations. These are (1) ROSAT All-Sky Survey data[27] and (2) XMM-Newton data[28] (observation identification number 556560101; principal investigator: P.J.W.).

**Data analysis and correction of systematics**

Because Lyman-α observations are photon-starved, the exposures are as long as the visibility of the target by HST allow. This yields spectra with exposure times of ~1,500 s to ~2,900 s, separated by ~40 min gaps. The data are acquired in time-tagged mode, which means that the arrival time of each photon hitting the detector is recorded by the far-ultraviolet multi-anode microchannel array (FUV-MAMA) of STIS. It is thus possible to split the exposures into several sub-exposures. We adjusted the sub-exposure time to obtain a good compromise between time sampling and signal-to-noise ratio. (The time sampling is important to assess systematic effects, as described later.) During visit 1, the total exposure time was 1,515 s (orbit 1) and 2,905 s (orbits 2 to 4). The first orbit in a visit always has less science time available before of the target acquisition. We used 4 × 380 s sub-exposures (orbit 1) and 5 × 581 sub-exposures (orbits 2 to 4). During visits 2 and 3, the total exposure time was 1,670 s (orbit 1) and 2,071 s (orbits 2 to 4). This was subdivided into 4 × 417 s sub-exposures (orbit 1) and 4 × 518 s sub-exposures (orbits 2 to 4), respectively. Extended Data Figs 3 to 5 show these time series, which are binned by a factor of 2 in Fig. 2 in the main text.

Variation of the temperatures in the optical telescope assembly and on the focal plan of HST during its revolution around the Earth is surmised to cause small changes of focus (for example, due to variation in the distance between the two M1 and M2 mirrors). For a given line of sight (hence for a given illumination pattern of the telescope), these variations depend on the HST orbital phase. They are consequently reproducible over several consecutive HST orbits, which means that their effect on the data, when present, can be corrected. This effect, known as the 'telescope breathing' effect, has been observed and corrected in STIS/CCD optical data[29] as well as in STIS/FUV-MAMA Lyman-α data[18].

We use the red part of the Lyman-α line, within the velocity range [+20,+250] km s$^{-1}$, as a probe for the systematic effects in general, and for the breathing effect in particular. In fact, the red peak and wing of the line are the only part of the spectrum unaffected (a priori) by the planet absorption and where there is enough flux to detect time-correlated effects. For each HST visit, we generate a raw light curve by integrating the Lyman-α flux over the velocity range given earlier, using the time-tag split exposures. These light curves are plotted in Extended Data Figs 3a, 4a and 5a, for visits 1, 2 and 3, respectively. If present in the data, the breathing effect would appear as a correlation in the light curves phase-folded over the HST orbital period (96 min). The phase-folded light curves for the three visits are shown in Extended Data Figs 3b, 4b and 5b. We then attempt to fit a polynomial to each phase-folded light curve and examine the dispersion of the residuals in the observed minus calculated data. When a significant correlation is found (as in visits 2 and 3), a phase-dependent correcting coefficient is applied to the whole spectrum.

For visit 1 data, we do not notice any significant correlation and find that a constant (0th-order polynomial) provides a better fit to the data ($\chi^2$ = 8.92 for 16 degrees of freedom with a Bayesian information criterion (BIC) of 17.04) than any higher-order polynomials. Consequently, we do not apply any breathing correction for this visit. For visit 2 data, we do find an improvement (over a constant) in the fit of the phase-folded light curve with straight lines ($\chi^2$ = 0.88 for 14 degrees of freedom, BIC = 7.12) and apply this correction to the raw data. For visit 3 data, we find a linear correlation in the phase-folded light curve. We correct the raw data with the best-fit straight lines ($\chi^2$ = 2.23 for 14 degrees of freedom, BIC = 12.7). The light curves corrected for breathing effect are shown in Extended Data Figs 4c and 5c for visits 2 and 3, respectively.

After having corrected for HST orbital phase-dependent effects, we now look at instrumental effects occurring on the timescale of a full HST visit (four HST orbits). Such effects typically include 'ramps' linked with long-term thermal relaxation of HST instruments. Our starting point is the Lyman-α red-wing light curves corrected for the breathing effect. Visit 1 still does not show any time correlation. In contrast, the visit 2 and 3 data show hints for a flux linearly decreasing with time. We chose to correct for this trend, by fitting a straight line to the data. This procedure yields a time-dependent correction coefficient for each visit. The final light curves corrected for (divided by) these trends are shown in Extended Data Figs 4d and 5d. The correction coefficients are applied to all spectra in visits 2 and 3. The data analysis and interpretation presented in the main text is based on these corrected spectra. These spectra are plotted for each visit in Extended Data Fig. 1. They clearly show that the signal is reproduced at each epoch, which strengthens our detection. In the main text, Fig. 1 presents these spectra averaged over visits 1, 2 and 3.

The amplitude of the correction of systematic effects performed as described earlier, either at the timescale of one HST orbit (breathing) or one HST visit (normalization), is small with respect to the absorption signal detected on the blue part of the Lyman-α line. The blue absorption would show up around the transit epoch almost unchanged, even in the absence of both corrections described earlier. Finally, the absorption signal by the exospheric cloud is so large that it can be measured directly in physical flux units. Our detection is thus independent of any normalization of the transit light curve.

**Determination of the velocity interval of the absorption signature**

In each individual spectrum of a given HST visit, we searched for the most significant (best signal-to-noise ratio) absorption signature. We scanned the velocity range [−250,+250] km s$^{-1}$, excluding the low signal-to-noise-ratio region between −30 km s$^{-1}$ and +20 km s$^{-1}$, for absorption signals extending over at least three contiguous elements of resolution. The signals are estimated from the difference between the spectrum and the reference out-of-transit spectrum for each visit. For visits 2 and 3, we simply used the data collected during the first HST orbits. We indifferently searched for absorption or emission signals, but found only significant (>3$\sigma$) absorption features. We found that the range [−120,−40] km s$^{-1}$ captures most of the absorption features detected during these two visits. For visit 1, where the first orbit is already absorbed (since it is in the pre-transit phase), we could not rely on contemporaneous out-of-transit data for reference. We thus adopted the range used for visits 2 and 3.

On the basis of the assumption that observations obtained during the first HST orbit of visit 1 are not absorbed, ref. 6 measured a 22.9 ± 3.9% (1$\sigma$) absorption signal, with respect to this first HST orbit, over the range of [−215,−17] km s$^{-1}$. We verified that we would obtain a similar result (~24%) by making the same assumptions as in ref. 6. However, as we show here, the first exposures obtained during visit 1 already exhibit strong absorption by the exospheric cloud, and choosing the out-of-transit references of visits 2 and 3 leads to a much deeper absorption signal being measured.

**Transit ephemeris**

Several ephemerides of GJ 436b transit have been published in the literature[6,26,30–32] and we list them in Extended Data Table 2. Visit 1 transit observation has been published previously[6], which reports the use of the following ephemeris values: $P$ = 2.643850(90) days and $T_0$ = 2,454,279.436714(15), for the period and mid-transit time, respectively. They attribute the period value to a previous study[30], although the values eventually reported in these two studies[6,30] are different, as can be seen in Extended Data Table 2.

We recalculate the transit phases of the visit 1 observation using one of the most recent available ephemerides[26]. After taking into account the ~6 min offset between the HST time values expressed in modified Julian days (MJD) and the barycentric Julian days (BJD), we obtained the phase values for the four HST orbits of the visit 1 observation (−1.74 h, −0.35 h, +1.25 h and +2.85 h). The obtained values are all ~42 min in advance of the phase values reported by ref. 6 (−1.05 h, +0.35 h, +1.95 h and +3.54 h). Trying to understand where this substantial difference came from, we realized that we would obtain similar transit phases as ref. 6, offset by ~6 min, when using the $T_0$ value of ref. 30 and the period given by ref. 6, which does not correspond to any value previously reported in the literature. The remaining ~6 min difference presumably comes from our accounting for the delay between BJD and MJD$_{UTC}$.

**New interpretation of a previous analysis of the visit 1 data**

The 42 min offset has two important consequences with respect to the visit 1 data interpretation by ref. 6. These authors indeed detected an absorption signature starting at the very end of the optical transit, with a maximum ultraviolet absorption depth of 22.9 ± 3.9% (1$\sigma$) occurring post-transit (~2 h after the end of the optical transit), according to their ephemeris. Shifting their data points 42 min earlier brings back the start of their absorption signal in-transit. The new phase of the first HST orbit in visit 1 is now −1.74 h before the optical mid-transit time, which corresponds to the pre-transit phases in visits 2 and 3. Since we observe a significant (3.4$\sigma$) absorption signal of 17.7 ± 5.2% in these visits, it means that the first HST orbit from visit 1, that ref. 6 used as an out-of-transit reference, is already partially absorbed by the exospheric cloud. In fact, the flux measured during the first HST orbits of visits 2 and 3 is significantly (3.4$\sigma$) higher than the flux measured in the second orbits of these visits (by the value of 17.7 ± 5.2% quoted earlier), and in good agreement with the flux measured ~30 h after transit, several years before in visit 0 (ref. 17), as can be seen in Fig. 2. The absorbed flux measured in the second HST orbits of visits 2 and 3 is a close match to the flux measured during the first HST orbit of visit 1. Using updated and more accurate transit ephemerides thus allows us to conciliate all HST observations within a common picture, quite different from the one painted by ref. 6. Not only does this substantially change the timing of the ultraviolet transit, but it also changes the amplitude of the measured absorption by a factor of ~2.5. This is critical for understanding the hydrogen atom dynamics and the exospheric cloud structure.

**X-ray observations and EUV flux of GJ 436**

Magnetic activity could induce variations in the stellar emission. We use the unabsorbed red part of the line to estimate that the intrinsic stellar variability at GJ 436 should not exceed 5% to 11% of the Lyman-$\alpha$ flux during one visit. Our Chandra X-ray data cover pre-transit phases at four epochs, two of which during HST visits 2 and 3, and we use them to monitor stellar activity. We detect a faint and soft X-ray source at the coordinates of GJ 436 in each of the four Chandra observations, with count rates of 2.0 ± 0.3, 3.1 ± 0.4, 2.4 ± 0.4 and 2.6 ± 0.4 ks$^{-1}$ for the February, April, June 2013 and June 2014 visits respectively. We extracted light curves (Extended Data Fig. 5) and spectra for each of these visits, as well as a combined spectrum for all four visits (Extended Data Fig. 6).

We also reanalysed the XMM-Newton EPIC-pn spectrum of GJ 436 (ref. 28) (Extended Data Fig. 6) and found a higher X-ray flux and luminosity than previously reported. We find an acceptable fit to the XMM-Newton spectrum with a two temperature thermal plasma model (APEC) with fitted temperatures of 0.09 and 0.38 keV and abundances of 0.18 Solar (fixed to the value listed by ref. 28). We measure a flux of $4.6 \times 10^{-14}$ erg s$^{-1}$ cm$^{-2}$ in the band previously used[28] (0.12–2.48 keV or 0.5–10 nm), giving a luminosity in this band of $5.7 \times 10^{26}$ erg s$^{-1}$ (ref. 28 values were $7.4 \times 10^{-15}$ erg s$^{-1}$ cm$^{-2}$ and $9.1 \times 10^{25}$ erg s$^{-1}$, respectively). Our higher values for flux and luminosity are in much closer agreement with the measurement made from the ROSAT All-Sky Survey data for GJ 436 (ref. 27). We also inspected the ROSAT data and confirmed the marginal detection of the source with approximately eight photons, all below 0.3 keV.

We fitted the combined spectrum from our four Chandra observations simultaneously with the XMM-Newton EPIC-pn spectrum (Extended Data Fig. 6). The Chandra and XMM-Newton spectra match well and we find an acceptable simultaneous fit with the two-temperature model described earlier. The temperatures and relative normalization of the two components were left free but forced to take the same values for the two data sets. We find temperatures of 0.097 keV and 0.72 keV, or approximately 1 MK and 8 MK. Only the fluxes were allowed to vary between the XMM-Newton and Chandra spectra, and this combined fit can be seen in Extended Data Fig. 6 (XMM-Newton PN is black, Chandra ACIS-S is red). In the overlapping energy range (0.234–2.0 keV) X-ray fluxes are $1.84 \times 10^{-14}$ erg s$^{-1}$ cm$^{-2}$ for XMM-Newton (December 2008) and $1.97 \times 10^{-14}$ erg s$^{-1}$ cm$^{-2}$ for Chandra (averaged across the four observations in 2013 and 2014).

We also fitted the four Chandra spectra individually, fixing the two temperatures to those derived from the combined fit to the XMM-Newton and Chandra data. This allowed us to determine the emission measures of the two temperature components as a function of time, which are plotted in Extended Data Fig. 7. We find that the higher temperature component, which dominates the spectrum above 0.5 keV, is more variable than the lower temperature component (dominating below 0.5 keV), as shown by Extended Data Fig. 8. The variability of the higher temperature component accounts for most of the count rate variability in the Chandra observations and is presumably caused by stellar flares. The relatively stable lower temperature component is a useful proxy for the unobserved EUV emission that dominates the high-energy irradiation of the planet and probably drives the mass loss. The final two Chandra observations are contemporaneous with our HST visits 2 and 3, and while the emission measure of the high temperature component increases by a factor 2.3 between these observations, the emission measure of the low temperature component increases by only 7%.

This helps to explain the very similar ultraviolet absorption depths and implied mass loss rates seen in HST visits 2 and 3. It can also be seen in Extended Data Fig. 8 that the emission measure of the lower temperature component was much lower during the first two Chandra observations in February and April 2013. HST visits at these times might have observed lower mass loss rates from the planet.

We use the X-ray flux measured in our third Chandra observation (contemporaneous with HST visit 2) to estimate the EUV flux irradiating the planet. The fitted model implies a flux in the 0.12–2.48 keV band (0.5–10 nm) of $4.7 \times 10^{-14}$ erg s$^{-1}$ cm$^{-2}$ (consistent with the flux at the time of the XMM-Newton observation). Converting to a flux at the surface of the star and comparing with the relation in figure 2 of ref. 33 leads to an estimate of the EUV flux at a reference separation of 1 au of 1.0 erg s$^{-1}$ cm$^{-2}$ (12.4–91.2 nm), which is a factor of 5 higher than the X-ray flux at 1 au of 0.21 erg s$^{-1}$ cm$^{-2}$ (0.5–10 nm).

This X-ray-derived EUV flux compares favourably with an estimate based on the strength of the Lyman-$\alpha$ line. Taking the total reconstructed line flux of $3.5 \times 10^{-13}$ erg s$^{-1}$ cm$^{-2}$ (ref. 24) and using the model from table 6 of ref. 34 gives an estimated EUV line flux at 1 au of 1.06 erg s$^{-1}$ cm$^{-2}$ (10–91.2 nm). The consistency of the EUV flux estimates based on our X-ray and Lyman-$\alpha$ observations gives confidence that the high energy irradiation of the planet is well constrained despite being unobservable directly.

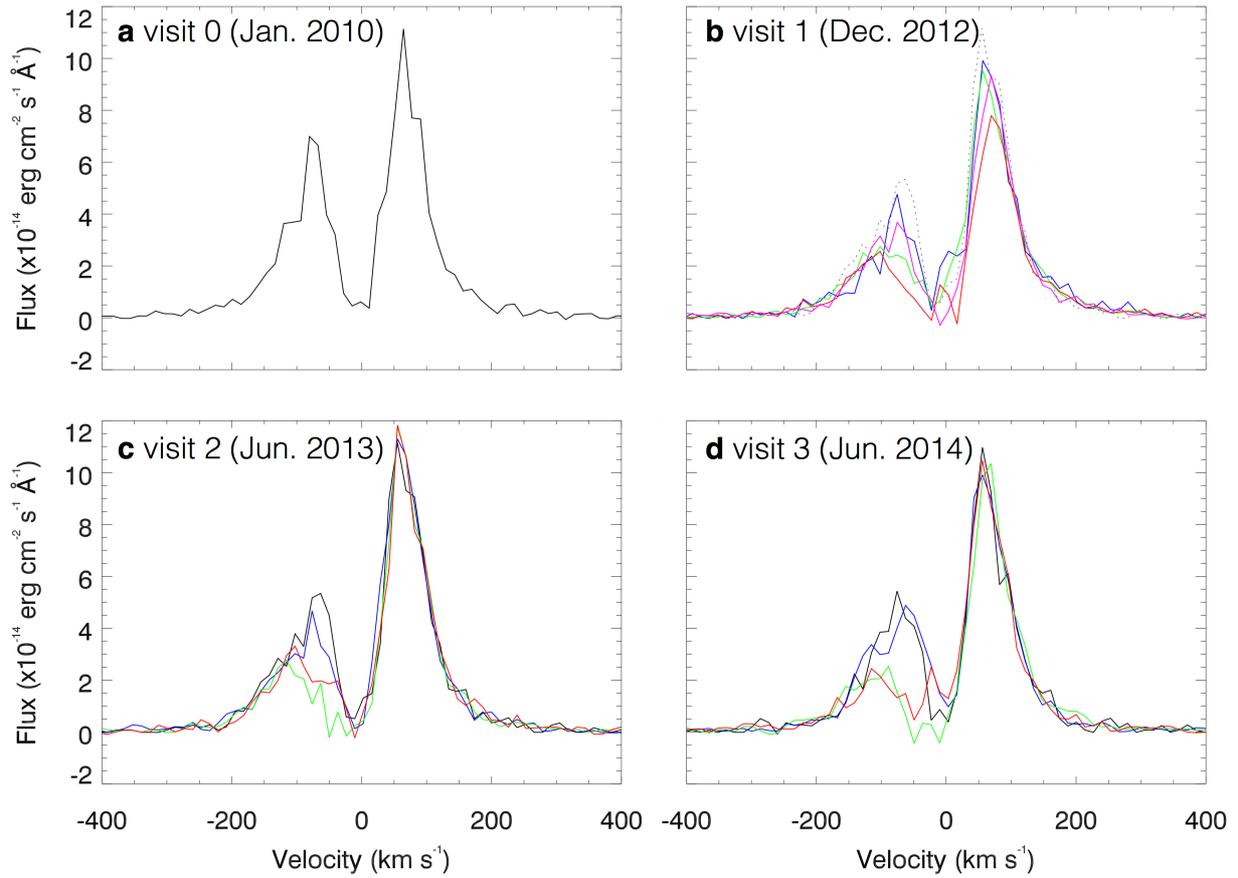

**Extended Data Figure 1 | Time evolution of GJ 436 Lyman-α line during each HST visit.** There is one spectrum per HST orbit. Colours indicate the phase with respect to the optical transit: out-of-transit (black), pre-transit (blue), in-transit (green) and post-transit (red and magenta for the last HST orbit in visit 1). For visit 1 there is no out-of-transit spectrum, hence we over-plotted the out-of-transit spectrum from visit 2 (dotted line).

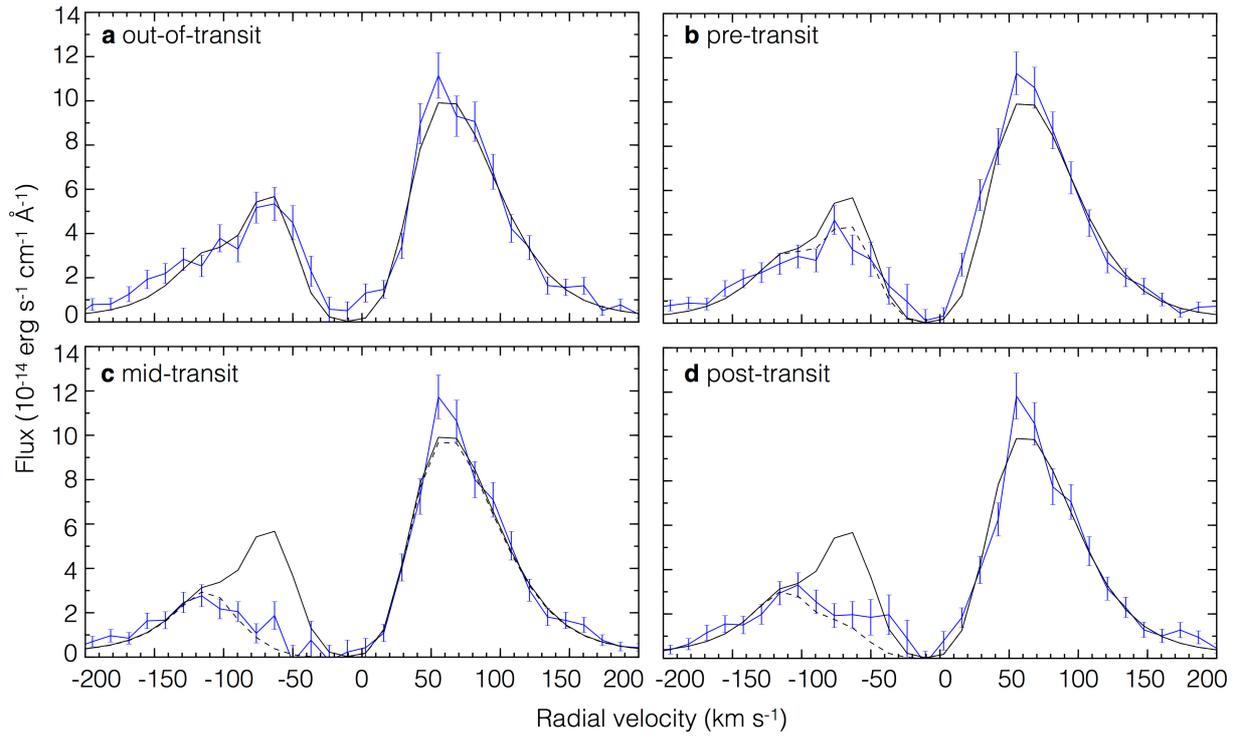

**Extended Data Figure 2 | Evolution of the Lyman-α line of GJ 436 reproduced with the numerical simulation. a–d**, The panels show the stellar emission line at different transit phases, here for visit 2 (all error bars are 1$\sigma$). **a**, The out-of-transit reference line (black curve) is the calculated profile best-fitting the observed out-of-transit line profile (blue curve), after taking into account the interstellar medium absorption and convolution by the instrumental line spread function. **b–d**, This theoretical profile is compared to observations (blue curves): the numerical simulation computes absorption in the theoretical profile and adjusts absorbed line profiles (dashed lines) as a function of time to the observations.

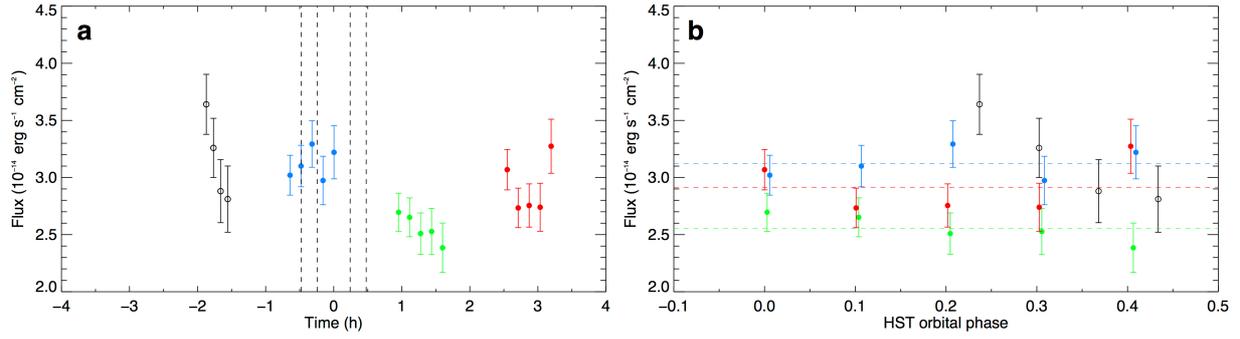

**Extended Data Figure 3 | Light curve from visit 1 data. a**, **b**, The light curve was calculated from integration of the flux over the red part of the line in the velocity range [+20,+250] km s$^{-1}$. The temporal (**a**) and phase-folded (**b**) light curves are shown. The different colours represent data acquired during different consecutive HST orbits. The vertical dashed lines in **a** indicate the location of the optical transit contacts. The horizontal dashed lines in **b** show the best-fit constants to each HST orbit (orbit 1 data—the open circles—are not fitted because of its different phasing). We did not apply any correction to this visit. We also did not find it necessary to trim out the first HST orbit in this visit (or in the subsequent ones) due to possible increased systematics, as described previously[18]. These time series have a higher sampling (by a factor of ~2) than those in Figure 2, which is made possible by exploiting the time-tag mode of data acquisition. The 1$\sigma$ error bars have been propagated accordingly.

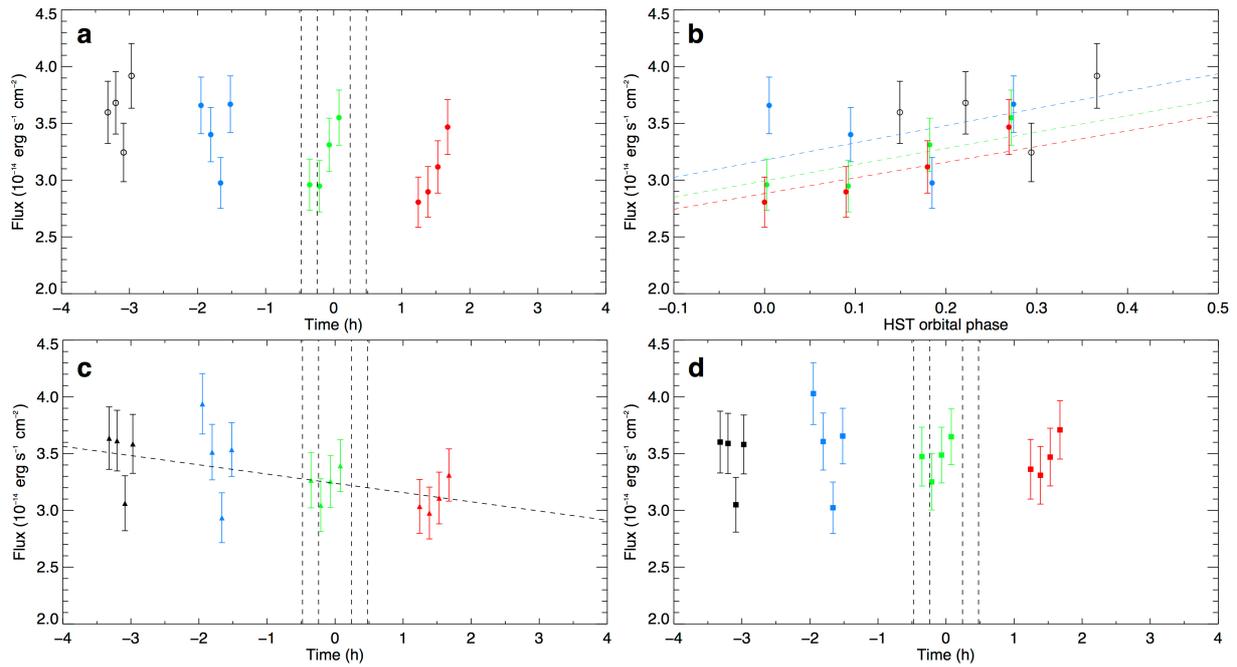

**Extended Data Figure 4 | Correction steps for the visit 2 data. a–d**, The different colours represent different consecutive HST orbits. The vertical dashed lines in the temporal light curves indicate the optical transit contact points. **a**, Raw light curve obtained from integration of the flux over the red part of the line in the velocity range $[+20,+250]$ km s$^{-1}$. **b**, Breathing correction. This is the same light curve as in **a**, phase-folded on the HST orbit. The dashed colour lines are linear fits to the different HST orbit data. Data from HST orbit 1 (empty black circles) are not taken into account because of their slightly different phasing. **c**, Light curve corrected from telescope breathing. A linear trend (dashed line) representing long-term systematics is fit to the data. **d**, Light curve corrected from telescope breathing and long-term systematics. All error bars are $1\sigma$.

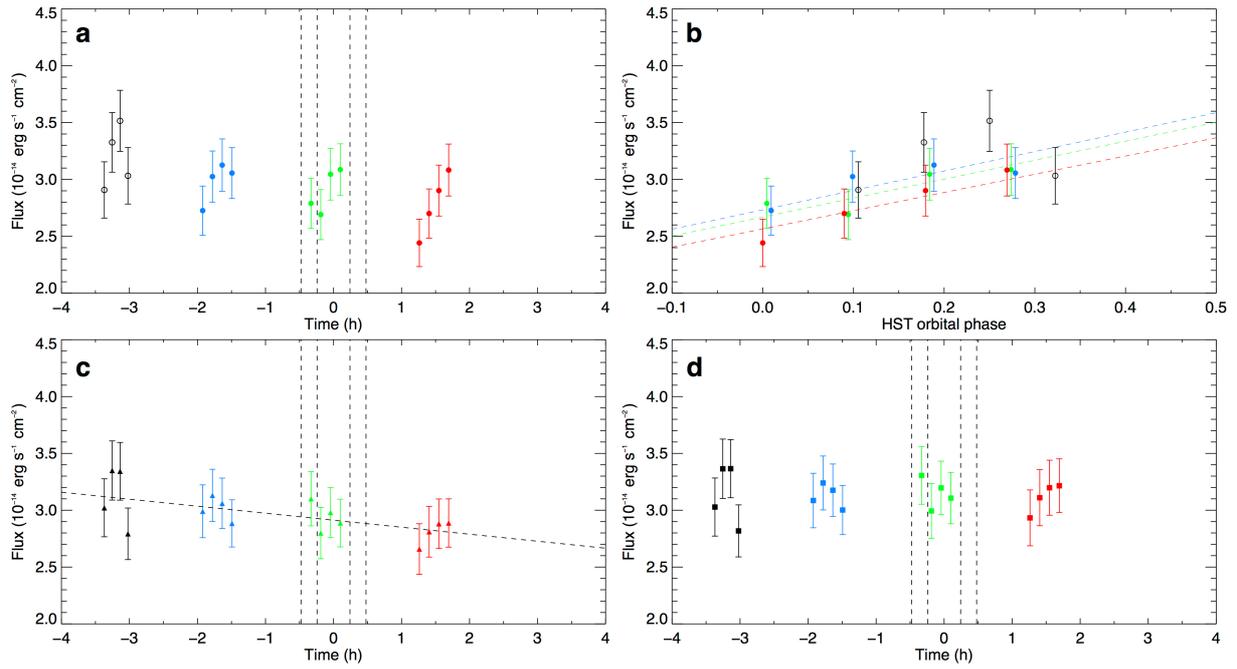

**Extended Data Figure 5 | Correction steps for the visit 3 data.** Identical to the description of Extended Data Fig. 2 for visit 2 data. All error bars are 1σ.

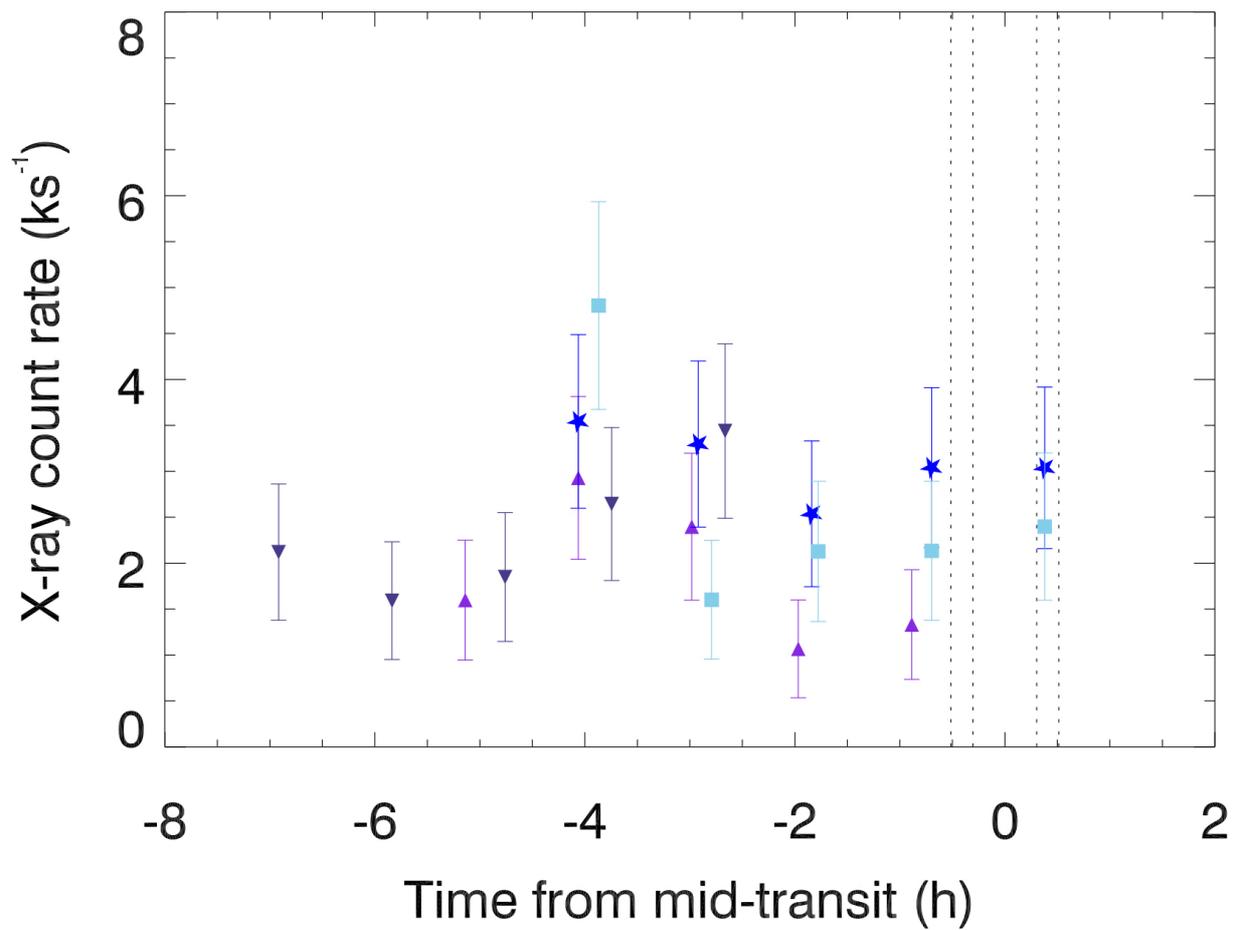

**Extended Data Figure 6 | Chandra X-ray counts of GJ 436.** The different symbols show the different visits. The June 2013 (stars) and June 2014 (squares) observations were contemporaneous with HST visits 2 and 3. The vertical dashed lines represent the contacts of the optical transit. All error bars are 1$\sigma$.

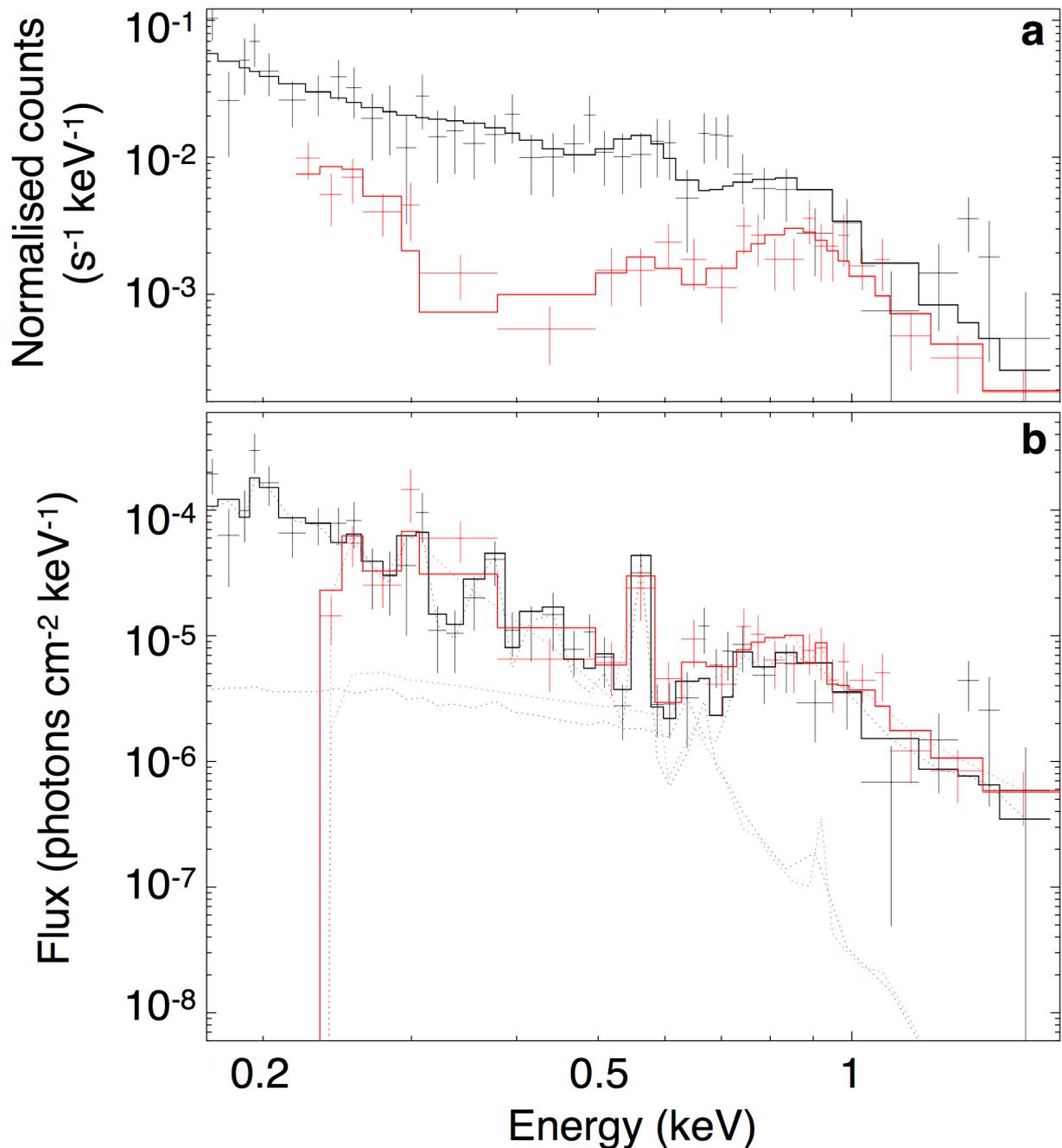

**Extended Data Figure 7 | X-ray spectrum of GJ 436 fitted with a two-temperature model. a, b,** The XMM-Newton EPIC-pn spectrum is shown in black, the combined Chandra ACIS-S spectrum is shown in red. **a,** Spectra in units of measured counts, with the Chandra spectrum below XMM-Newton because of the lower sensitivity of the instrument. **b,** Unfolded spectra in flux units; it can be seen that the XMM-Newton and Chandra spectra are consistent with each other. The dotted lines show the contributions of the individual temperature components. All error bars are $1\sigma$.

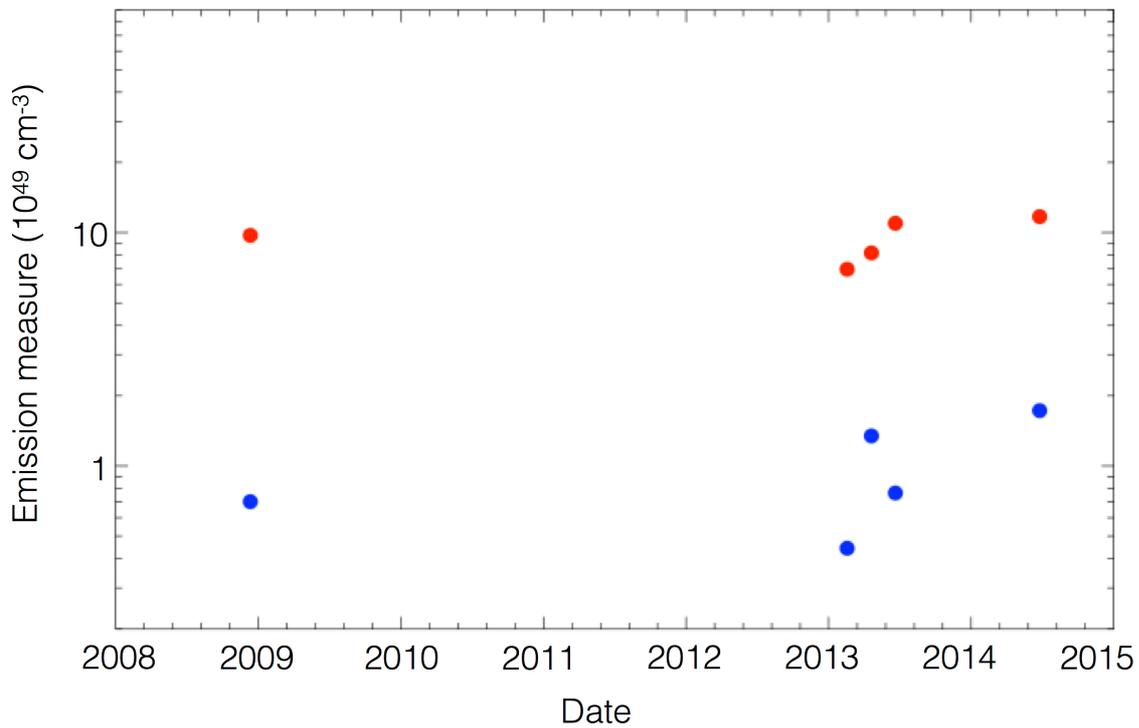

**Extended Data Figure 8 | Emission measures for the two temperature components fitted to the individual XMM-Newton and Chandra spectra.** The red points correspond to the low temperature component (0.097 keV) that dominates the spectra below 0.5 keV. The blue points correspond to the higher temperature component (0.72 keV) that dominates above 0.5 keV. The higher temperature component is more variable, probably due to the effect of stellar flares. The less variable lower temperature component is a proxy for the unobserved extreme-ultraviolet emission that dominates the high-energy irradiation of the planet. The last two points are contemporaneous with HST visits 2 and 3.

**Extended Data Table 1 | Log of the HST and Chandra observations**

| Observation date | HST/STIS/G140M | Chandra/ACIS-S |
|---|---|---|
| 2010-01-05 | GO#11817<br>1 HST orbit ("visit 0")<br>Ref. 17 | |
| 2012-12-07 | GTO#12034<br>1 transit (4 HST orbits, "visit 1")<br>Ref. 6 | |
| 2013-02-16 | GO#12965<br>Delayed, no data | Obs ID 14459<br>1 transit (18.77 ks) |
| 2013-04-18 | GO#12965<br>Target pointing problem, no data | Obs ID 15537<br>1 transit (19.75 ks) |
| 2013-06-18 | GO#12965<br>1 transit (4 HST orbits, "visit 2") | Obs ID 15536<br>1 transit (18.91 ks) |
| 2014-06-23 | GO#12965<br>1 transit (4 HST orbits, "visit 3") | Obs ID 15642<br>1 transit (18.77 ks) |

**Extended Data Table 2 | Ephemerides of GJ 436b transit**

| Mid-transit time $T_0$ (JD) | Period $P$ (days) | Reference |
|---|---|---|
| BJD 2,456,295.431924(45) | 2.643897 82(8) | Ref. 31 |
| BJD 2,454,865.084034(35) | 2.643898 03(30) | Ref. 32 |
| BJD 2,454,865.083208(42)* | 2.643897 9(3)* | Ref. 26 |
| HJD 2,454,415.62074(8) | 2.643890 4 | Ref. 30 |
| HJD 2,454,415.62074(8) | 2.643850(90) | Ref. 6 |

The offset between BJD and heliocentric Julian days (HJD) is small, typically ±4 s.
*Values used in the present work.